\documentclass[usenatbib]{mn2e}
\usepackage{natbib,epsfig,color}

\catcode`\@=11
\def\gta{\ifmmode{\,\mathrel{\mathpalette\@versim>\,}}
    \else{$\,\mathrel{\mathpalette\@versim>}\,$}\fi}
\def\lta{\ifmmode{\,\mathrel{\mathpalette\@versim<\,}}
    \else{$\,\mathrel{\mathpalette\@versim<}\,$}\fi}
\def\@versim#1#2{\lower 2.9truept \vbox{\baselineskip 0pt \lineskip
    0.5truept \ialign{$\m@th#1\hfil##\hfil$\crcr#2\crcr\sim\crcr}}}
\catcode`\@=12  

\usepackage{graphics} \usepackage{amssymb}
\usepackage{setspace} 

\newcommand{\beq}{\begin{equation}}
\newcommand{\eeq}{\end{equation}}
\newcommand{\eqref}[1]{eq.~(\ref{#1})}

\newcommand{\tabref}[1]{Table \ref{#1}}
\newcommand {\hi} {{\rm H}\,{\small\rm I}}

\newcommand {\Tcor}{T_{\rm cor}}
\newcommand {\Zcor}{Z_{\rm cor}}
\newcommand {\Tcl}{T_{\rm cl}}
\newcommand {\Zcl}{Z_{\rm cl}}
\newcommand {\vs}{v_{\rm s}}
\newcommand {\Rs}{R_{\rm s}}

\newcommand{\comments}[1]{#1}

\def\K{\,{\rm K}}
\def\Myr{\,{\rm Myr}}

\def\kpc{\,{\rm kpc}}\def\pc{\,{\rm pc}}
\def\kms{\,{\rm km\,s}^{-1}}
\def\cm{\,{\rm cm}}

\def\hi{H{\sc\,i}}

\def\HVC{{\sc hvc}}
\def\IVC{{\sc ivc}}

\def\CIV{C{\sc\,iv}}
\def\OV{O{\sc\,v}}
\def\OVI{O{\sc\,vi}}
\def\OVII{O{\sc\,vii}}
\def\ECHO{{\sc echo}}  

\defcitealias{FraternaliB08}{FB08}
\defcitealias{Marinacci10}{M10}

\newcommand {\gsim}{\,\lower.7ex\hbox{$\;\stackrel{\textstyle>}{\sim}\;$}}
\newcommand {\lsim}{\,\lower.7ex\hbox{$\;\stackrel{\textstyle<}{\sim}\;$}}
\newcommand{\coronae}{coronae}

\title[Rotation of galactic \coronae]
{Galactic fountains and the rotation of disc-galaxy \coronae}

\author[F. Marinacci et al.]
{Federico Marinacci$^{1}$\thanks{e-mail: federico.marinacci2@unibo.it},  Filippo Fraternali$^1$, Carlo Nipoti$^1$,
James Binney$^{2}$, \newauthor Luca Ciotti$^1$ and Pasquale Londrillo$^3$\\
$^1$Astronomy Department, University of Bologna, via Ranzani 1,
I-40127 Bologna, Italy\\ 
$^{2}$Rudolf Peierls Centre for Theoretical Physics, Oxford University, Keble
Road, Oxford OX1 3NP, UK\\
$^3$INAF-Bologna Astronomical Observatory, via Ranzani 1, I-40127 Bologna, Italy}

\begin{document}

\date{Accepted 2011 March 25. Received 2011 February 18}

\pagerange{\pageref{firstpage}--\pageref{lastpage}}
\pubyear{2011}
\maketitle

\label{firstpage}

\begin{abstract}
  In galaxies like the Milky Way, cold ($\sim 10^4 \K$) gas ejected
  from the disc by stellar activity (the so-called galactic-fountain
  gas) is expected to interact with the virial-temperature ($\sim 10^6
  \K$) gas of the corona. The associated transfer of momentum between
  cold and hot gas has important consequences for the
  dynamics of both gas phases. We quantify the effects of such an
  interaction using hydrodynamical simulations of cold clouds
  travelling through a hot medium at different relative velocities.
  Our main finding is that there is a velocity threshold between clouds
  and corona, of about $75\kms$, below which the hot gas ceases to
  absorb momentum from the cold clouds. It follows that in a disc
  galaxy like the Milky Way a static corona would be rapidly
  accelerated: the corona is expected to rotate and to lag, in the inner regions,
  by $\sim 80-120 \kms$ with respect to the cold disc.
  We also show how the existence of this velocity threshold can 
  explain the observed kinematics of the cold extra--planar gas.
\end{abstract}


\section{Introduction}\label{sect:intro}

Current cosmological models suggest that galaxies are embedded in
extended virial-temperature atmospheres called \coronae. These
\coronae\ are thought to contain a significant fraction (more than
$50\%$) of the baryons in the Universe on the basis of a combination
of big-bang nucleosynthesis theory and observations of the cosmic
microwave background
\citep[e.g.][]{Fukugita98,Fukugita04,Komatsu09}. In clusters of
galaxies the baryons associated with the hot intra-cluster medium are
directly detected in X-rays through their free-free emission
\citep{Sarazin09}, whilst in lower density environments, such as the
Local Group, the hot medium must be characterised by a much lower
X-ray surface brightness, below the sensitivity of the current
instrumentation \citep{Anderson10}.

The existence of coronal gas around our Galaxy was hypothesised by
\cite{Spitzer56} to provide the necessary pressure confinement to cold
clouds of interstellar gas that lie far from the Galactic plane. 
With the advent of UV spectroscopy, highly ionised material 
(in particular \CIV,\OV\
and \OVI) was detected along lines of sight to distant UV sources that
pass close to high- and intermediate-velocity clouds \citep[\HVC s and
\IVC s, see][]{Sembach03,Fox06}.  These clouds are not massive enough
to be confined by self-gravity and the detected species are believed
to lie at the interface between the coronal gas with $T \gsim 10^6\K$,
and the \HVC s/\IVC s, which are composed by cooler and partly neutral
gas.

In star-forming disc galaxies, high-sensitivity \hi\ observations have
revealed that roughly $10\%$ of the neutral gas is outside the thin
disc, at distances of a few kiloparsecs \citep[and references
  therein]{Oosterloo07, Fraternali09}.  The origin of this
extra--planar gas is not completely understood but it is now widely
accepted that a large fraction of it is produced by supernova-powered
bubbles which drive neutral and partly ionised gas out of the galactic
disc.  This gas is expected to travel through the halo and eventually
to fall back to the disc in a time-scale of $\sim 80-100\Myr$, a
mechanism which is known as galactic fountain
\citep{Shapiro76,Bregman80,HouckB90}.  Thus, in the regions within a
few kiloparsecs from the galactic disc a large number of galactic-fountain
``cold'' clouds (\IVC s in the Milky Way) should be travelling through
the ubiquitous coronal gas and an interaction between the two phases
is inevitable, with important consequences for the kinematics of both.

Unfortunately, the kinematics of the coronal gas is virtually
unconstrained: as mentioned, most of the evidence for this gas is
indirect, and from the available data it is not possible to extract
reliable kinematic information. A potentially useful direct detection
comes from the observation of \OVII\ absorption lines (which probe
material at temperatures around $10^6 \K$) towards bright quasars
\citep[e.g.][]{Rasmussen03}.  However due to the poor spectral
resolution of the current X-ray instruments, these observations are
impractical \citep{Bregman07,Yao08}.

On the other hand, the kinematics of the extra--planar cold gas has
been studied extensively with \hi\ and optical line data
\citep{Swaters97, Fraternali02, Heald07, Oosterloo07, Kamphuis08}.  It
was found that the extra--planar gas rotates with a speed that
decreases steadily with increasing distance from the mid-plane of the
galaxy.  A number of models have been proposed to explain this vertical velocity
gradient (e.g. \citeauthor{Barnabe06} \citeyear{Barnabe06};
\citeauthor{Kaufmann06} \citeyear{Kaufmann06};
\citeauthor{Melioli09} \citeyear{Melioli09};
\citeauthor{Struck09} \citeyear{Struck09};
\citeauthor{Marinacci10b} 2010b).
In particular, it has been shown that the
\textit{rotational} gradient is larger than what would be expected for
a collection of galactic-fountain clouds that travel through the halo
region on pure ballistic trajectories \citep{FraternaliB06}, so the
natural next step is to consider how clouds' trajectories are affected
by interactions with the ambient corona.

\citeauthor{FraternaliB08} \citepalias[\citeyear{FraternaliB08},
hereafter][]{FraternaliB08} considered two types of cloud--corona
interaction: drag and accretion.  The drag between the \hi\ clouds and a
supposedly lagging corona was proposed, among others by \cite{Collins02} and
\cite{Barnabe06}, to explain the negative vertical rotation gradient of the
extra--planar gas.  However, \citetalias{FraternaliB08} showed that this
mechanism is not viable because of the resulting rapid acceleration of the corona, which
causes the drag to vanish in less than a cloud orbital time.  Thus, if the
drag is as efficient as one would expect from purely mechanical arguments,
the corona should rotate at a speed comparable to that
of the disc, with a consequent vanishing of the drag.
\citetalias{FraternaliB08} then investigated the possibility that \hi\ clouds
accrete some of the ambient gas that lies in their paths.
They showed that if the ambient gas has a
relatively low angular momentum about the galaxy's spin axis, its accretion
 onto \hi\ clouds could explain the observed dynamics of
the fountain clouds. 

In a preliminary study, \citeauthor{Marinacci10} \citepalias[2010a,
  hereafter][]{Marinacci10} presented hydrodynamical simulations of
cold clouds travelling through a hot medium with properties similar to the gas
of a galactic corona. They found that radiative cooling can be
efficient in the turbulent wakes of the clouds, allowing coronal
material to condense in these wakes.  Therefore, even though the
original cloud is ablated, there is a net transfer of mass from the
corona to the whole body of cool gas, supporting the
\citetalias{FraternaliB08} framework.

In this paper we extend the study of the interaction between the
clouds and the corona presented by \citetalias{Marinacci10} in two
ways: (i) we use an upgraded version of the hydro-code taking into
account the variation in the mean molecular weight of the gas at
$T\simeq10^4\K$ (mainly due to hydrogen recombination) and, more
importantly, the different metallicity of the disc and of the coronal
gas; (ii) we study a much wider range of relative cloud--corona velocities.
Our major finding is that there is a velocity threshold between cloud
and corona below which the corona does not absorb momentum from the
cloud. We argue that the corona must be spinning fast enough for most
clouds to move through the corona at a speed smaller than this
velocity threshold. This prediction will be tested by future soft
X-ray observations.

The paper is organised as follows:  Section 
\ref{sec:simuls} explains which simulations were run, and the results of
these simulations are analysed in Section \ref{sect:results}; in Section
\ref{sect:discussion} we discuss the implications of our
results for the kinematics of both the corona and the clouds,
while Section \ref{sect:conclusions} sums up.

\begin{table}
\centering
\begin{tabular}{cccc}
\hline\hline
Simulation & $v_0$ & $n_{\rm cor}$ &  $M_{\rm cl}$ \\
&  $(\kms)$ & $({\rm cm}^{-3})$ & $(10^4~M_{\sun})$ \\
\hline
\multicolumn{4}{c}{Standard simulations}\\
\hline
v200 & 200  & $10^{-3}$ & $2.4$ \\
v150 & 150 & $10^{-3}$ & $2.4$ \\
v100 & 100 & $10^{-3}$ & $2.4$ \\
v75 & 75 & $10^{-3}$ & $2.4$ \\
\hline
\multicolumn{4}{c}{High-density simulations}\\
\hline
v100--h & 100 & $2\times10^{-3}$ & $4.8$ \\
v75--h & 75 & $2\times10^{-3}$ & $4.8$ \\
\hline
\multicolumn{4}{c}{Low-density simulations}\\
\hline
v100--l & 100 & $5\times10^{-4}$ & $1.2$\\
v75--l & 75 & $5\times10^{-4}$ & $1.2$ \\
v50--l & 50 & $5\times10^{-4}$ & $1.2$\\
\hline\hline
\end{tabular}
\caption{Parameters of the simulations.  The initial temperature and
  metallicity of the corona are $\Tcor = 2\times10^6 \K$ and $\Zcor =
  0.1~Z_{\sun}$, respectively, while for the cloud $\Tcl = 10^4 \K$
  and $\Zcl = Z_{\sun}$. All models are evolved up to $60 \Myr$.  The
  grid size is 6114x1024 with spatial resolution $\sim 2\times2~\pc$
  for all the simulations.  In the case of v100 an additional run,
  with the same spatial resolution and grid size 6144x2048, was
  performed. }
\label{tab:simulations}
\end{table}

\section{Set up of the hydrodynamical simulations}
\label{sec:simuls} 

The initial conditions of the simulations are similar to those of
\citetalias{Marinacci10}: we simulate a cloud at $T = 10^4\K$ with
radius $R_{\rm cl} = 100 \pc$ that moves at a speed $v_0$ through a
stationary and homogeneous background at $T = 2\times 10^6 \K$, which
represents the corona.  While in \citetalias{Marinacci10} the \textit{relative}
speed between the cloud and the corona was fixed at $v_0=75 \kms$,
here we explore different relative speeds in the range $50\leq v_0\leq
200 \kms$. If these relative speeds are interpreted as differences
in rotational velocities, then 
two extreme possibilities are encompassed by the simulations: (i) the
case of a non-rotating corona ($v_0=200 \kms$) and (ii) a corona which is
rotating with a speed close to that of the disc ($v_0=50 \kms$).  As
on ballistic trajectories the galactocentric radii of the clouds would
vary by less than $30\%$, and the clouds would reach heights above the
galactic plane of a few kiloparsecs \citep[see][]{FraternaliB06}, 
the expected density variations of the
coronal gas are not strong. Thus, as in \citetalias{Marinacci10}, we
simplify our treatment by assuming a homogeneous corona and neglecting
the effect of the gravitational field of the galaxy. We also neglect
the self gravity of the gas, because the cloud masses are much smaller
than their Jeans mass. At the beginning of each simulation the cloud is
in pressure equilibrium with its surroundings and the evolution of the
system is followed for $60 \Myr$ (significantly longer than the
$25\Myr$ spanned by most runs in \citetalias{Marinacci10}).  In our
standard simulations (see \tabref{tab:simulations}) the density of the
background is $n = 10^{-3}\,{\rm cm^{-3}}$.  For each value of $v_0$,
we ran two simulations: one (dissipative) in which the gas is allowed
to cool radiatively, another (adiabatic) without radiative
cooling. The more realistic cases are those in which the gas is
allowed to cool, but the adiabatic runs enable us to distinguish the
effect of cooling (strongly dependent on the density, temperature and
metallicity of the gas) from non-dissipative effects like mixing and
hydrodynamical instabilities.  Two additional sets of dissipative
simulations have been run with half (low-density simulations in
\tabref{tab:simulations}) or twice (high-density simulations in
\tabref{tab:simulations}) the coronal density to assess the influence
of this latter on the results.  In all initial conditions the
cloud's metallicity $\Zcl$ takes the solar value $Z_{\sun}$ and the
coronal metallicity is $\Zcor=0.1 Z_{\sun}$ \citep[see,
  e.g.][]{Sembach03,Shull09}. The metallicity of the fluid is allowed
to vary from the initial conditions: in this respect the
present simulations differ from those of \citetalias{Marinacci10}, in
which the metallicity was constant. This upgrade is important since
the metallicity of a cloud ejected from the disc and that of the
corona can differ by one order of magnitude or more, and the cooling
rate of the gas critically depends on its metal content. 
In any case, we stress that the
main results of \citetalias{Marinacci10} are confirmed by the present
investigation: for instance, the global behaviour of \citetalias{Marinacci10}
simulation with $n = 10^{-3} \cm^{-3}$ and {\it uniform} metallicity $Z =
0.3~Z_{\sun}$ is similar to ours with the same initial coronal
density and cloud speed, but $\Zcl=Z_{\sun}$ and
$\Zcor=0.1\,Z_{\sun}$.

The simulations are performed with \ECHO, a high-order,
shock-capturing Eulerian hydrodynamical code: a detailed description
and tests of it can be found in \cite{DelZanna02} and
\cite{DelZanna07}.  In the current version of the code (\ECHO ++) not only the
metallicity evolves, but also the fluid's molecular weight
$\mu$ is a function of temperature. The variation of $Z$ and $\mu$ is
accounted for in the implementation of radiative cooling, which is
modelled according to the prescription of \cite{Sutherland93}. The
calculations are performed on a two-dimensional Cartesian grid with
open boundary conditions imposed to all the sides of the computational
domain. Therefore, one of the
dimensions perpendicular to the cloud's velocity has been suppressed,
and we are in effect simulating a flow around an infinite cylindrical
cloud. The cylinder is moving perpendicular to its long axis and has
initially a circular cross-section of radius $R_{\rm cl}$. From the
simulations we obtain quantities per unit length of the cylinder and
we relate these to the corresponding quantities for an initially
spherical cloud of radius $R_{\rm cl}$ by multiplying the cylindrical
results by the length ${4\over3}R_{\rm cl}$ within which the mass of
the cylinder equals the mass of the spherical cloud.  This correction
is included in the values of the cloud mass $M_{\rm cl}$ quoted in
\tabref{tab:simulations}. We note that, due to pressure equilibrium
in the initial conditions, for the same gas number
density, temperature and cloud size $M_{\rm cl}$ is higher here than
in \citetalias{Marinacci10}, because here we account for the fact that
$\mu$ depends on temperature, while in \citetalias{Marinacci10} $\mu$
is fixed at the coronal value.

\begin{figure*}
\centerline{\epsfig{file=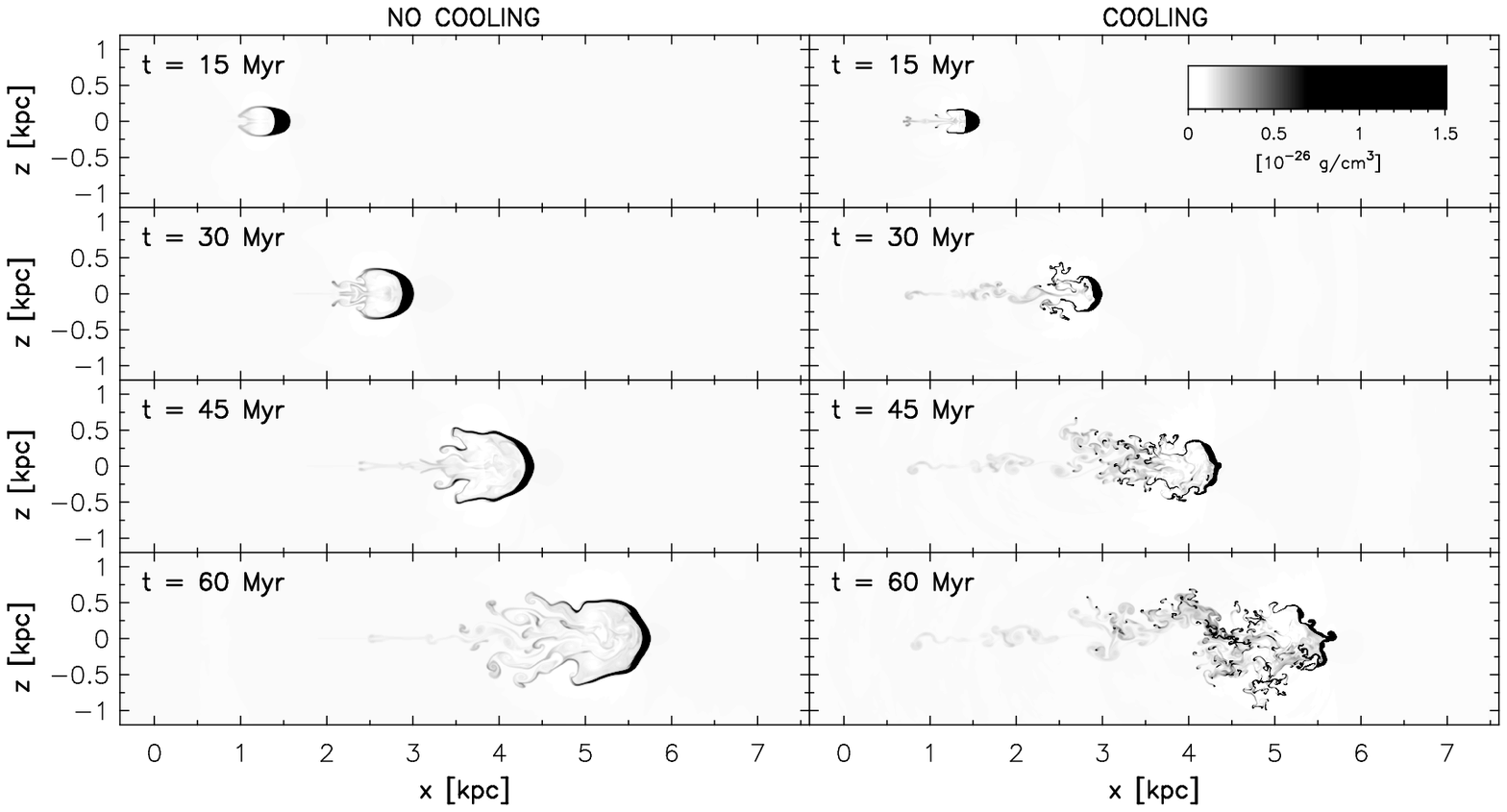,width=\hsize,viewport=35 480
555 742}}
\caption{Density snapshots of the adiabatic (left panels) and
  dissipative (right panels) standard simulations with $v_0=100
  \kms$ (v100 in Table~\ref{tab:simulations}). The time at which
  the snapshot has been taken is indicated in each panel. The initial
  position of the cloud centre is $x=0$.}
\label{fig:simulmom}
\end{figure*}
\begin{figure*}
\centerline{\epsfig{file=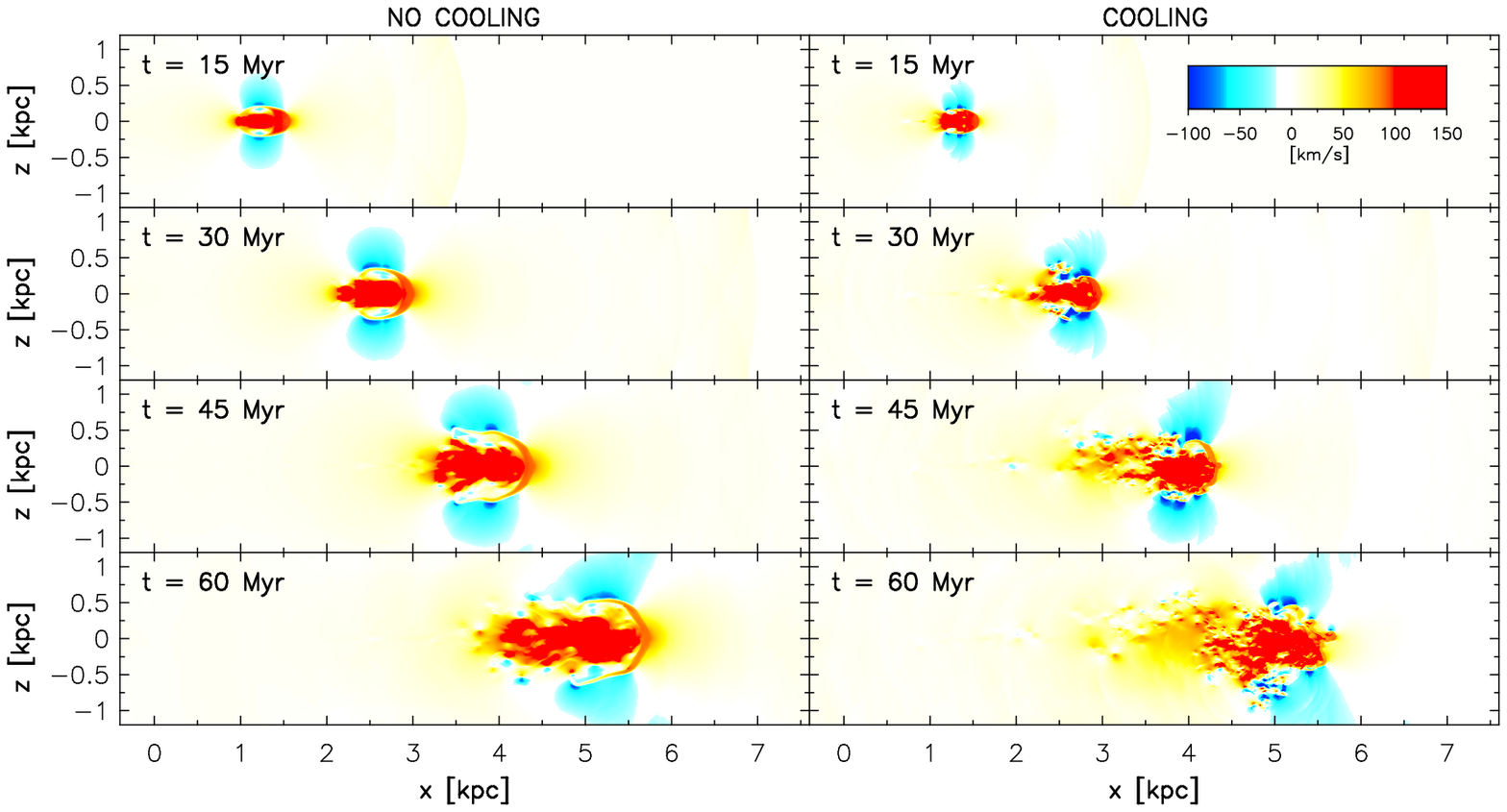,width=\hsize,viewport=35 480 555
742}}
\caption{Same as Fig.~\ref{fig:simulmom}, but plotting
the velocity along the cloud's direction of motion ($v_x$).}
\label{fig:simulmom2}
\end{figure*}

\section{Results of the simulations}\label{sect:results}

We now describe the results of the hydrodynamical simulations, with
emphasis on the mass and momentum transfer between the cloud and the
ambient gas. While the two phases are unambiguously separated in the
initial conditions (the cloud gas is at $T=10^4\K$ and the coronal gas is at
$T=2\times 10^6 \K$), at later times, as a consequence of turbulent
mixing and radiative cooling, we expect to have a significant amount
of gas at intermediate temperatures.  Therefore, at all times in the
simulations, we define ``hot'' the gas at $T>10^6 \K$ and ``cold'' the
gas at $T < 3\times 10^4 \K$, identifying the hot phase with the
corona and the cold phase with the cloud.  We verified that our
results do not depend significantly on the specific choice of these
temperature cuts.

\subsection{Interaction between the cloud and the ambient medium} \label{sect:morph_cloud}

Qualitatively, the evolution of the cloud in the present simulations
is similar to that observed in the simulations of
\citetalias{Marinacci10}. Due to the ram pressure arising from the
motion, the cloud feels a drag which causes it to decelerate while the
background gains momentum. The body of the cloud is squashed, and
behind it a wide turbulent wake forms, in which the coronal and cloud
gas mix efficiently and radiative cooling is particularly effective,
with the consequence that the cloud mass tend to increase with time.
The wake has a complex velocity structure, with some fluid elements
moving faster and other slower than the body of the cloud.

This behaviour is apparent from Figs.~\ref{fig:simulmom} and
\ref{fig:simulmom2}, showing, respectively, density and velocity
snapshots, at four different times, of the dissipative standard
simulation with $v_0 = 100 \kms$ (right panels) and, for comparison,
of the corresponding adiabatic simulation (left panels).  The diagrams
in the bottom panels clearly show that the cloud is decelerated by
drag: at $100 \kms$ $\simeq 0.1\, \kpc\Myr^{-1}$ the cloud would
travel, with no drag, a distance of about $6\kpc$ in $60 \Myr$, but
neither in the presence nor in the absence of radiative cooling does
the main body of the cloud reach that distance.  The structure of the
body of the cloud and of the wake is strongly influenced by the
presence of radiative cooling: the wake is more elongated and less
laterally extended (except perhaps in the $60 \Myr$ panel) when the
radiative cooling of the gas is permitted.  The extent of the velocity
structure of the wake is greater, both laterally and in elongation, in
the dissipative than in the adiabatic case at late times (see
Fig.~\ref{fig:simulmom2}). Just behind the cloud there is a region in
which the velocities reach their maximum values and this region is
more compact (less elongated) when cooling occurs. In front of the
cloud the coronal gas is pushed by the cloud and therefore also in
this region the gas has positive velocity.  However, in contrast to
what happens behind the main body of the cloud, this region is more
extended in the absence of cooling.

\begin{figure} 
\centerline{\epsfig{file=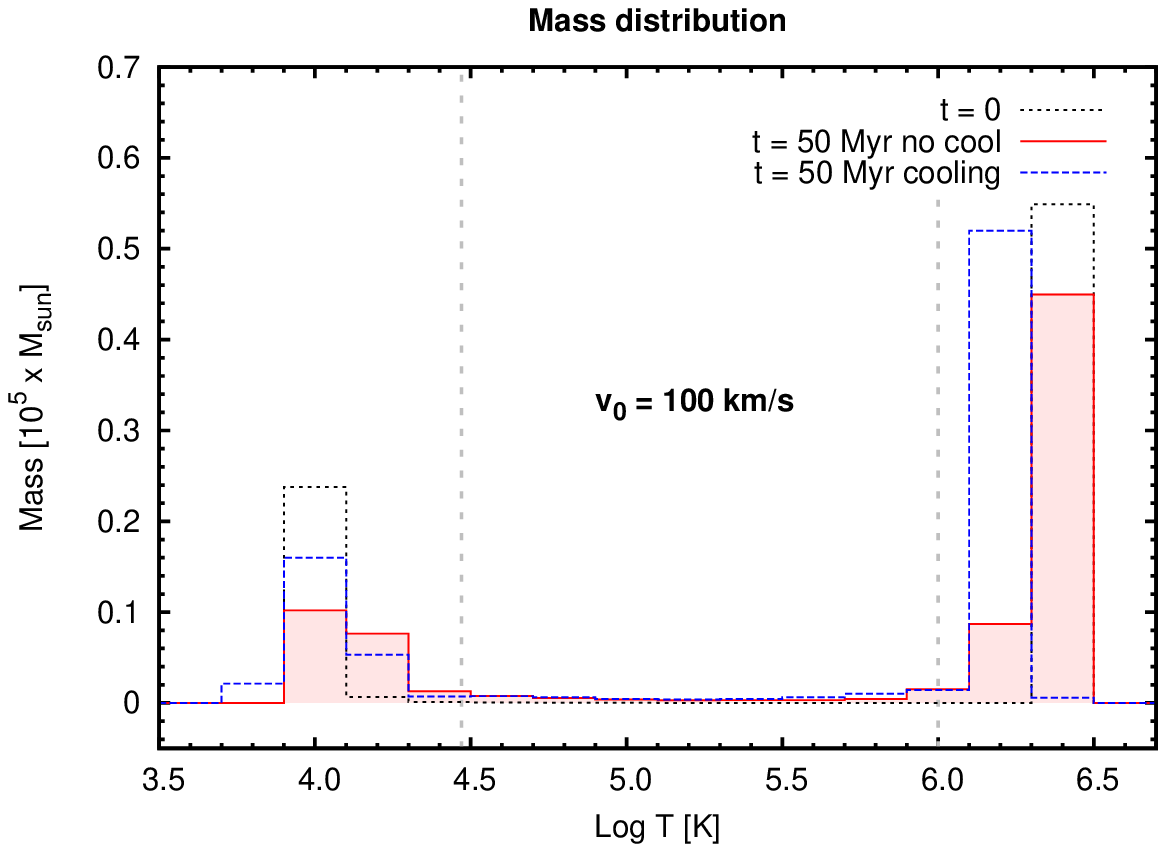,width=\hsize}}
\centerline{\epsfig{file=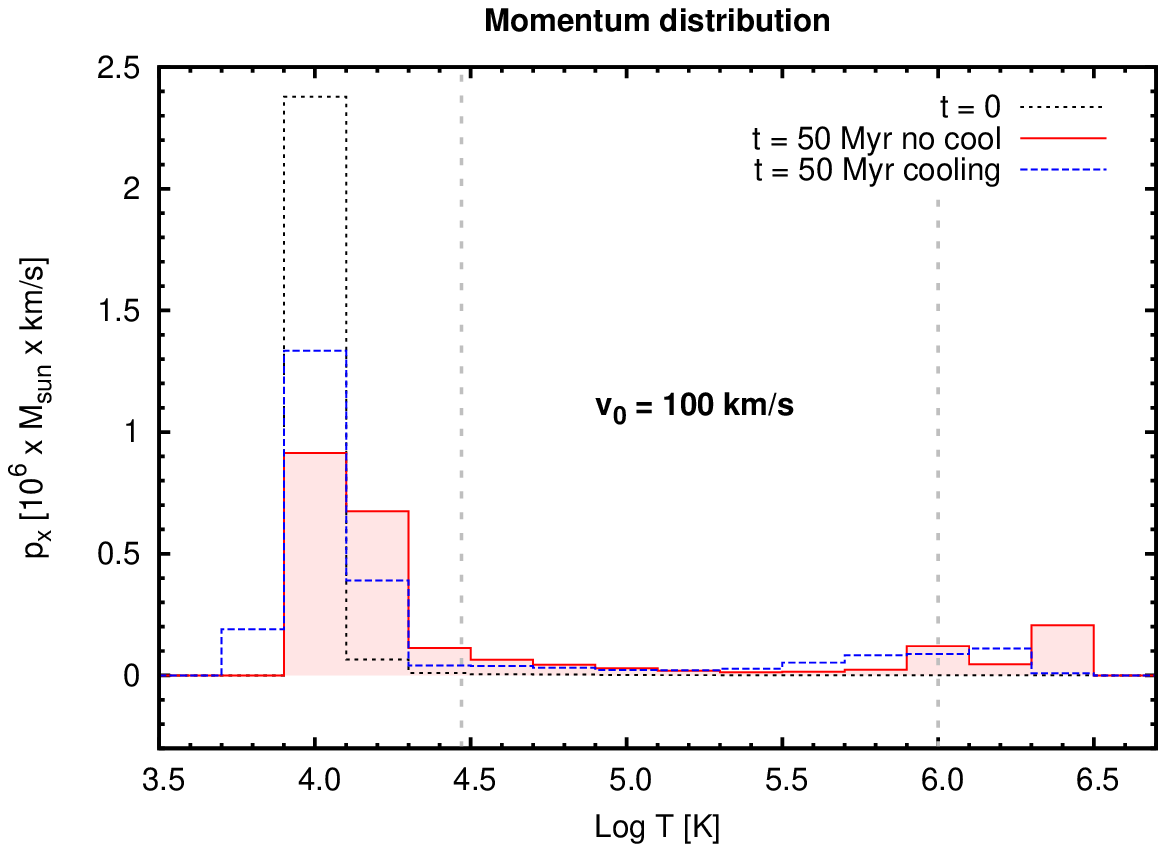,width=\hsize}}
\caption{Distributions of mass (upper panel) and momentum (lower
  panel) as functions of the gas temperature after 50$\Myr$ of
  evolution of the the adiabatic (shaded) and dissipative (long-dashed
  line) standard simulations with $v_0 = 100 \kms$; the black
  short-dashed line is the initial mass or momentum distribution. The
  vertical grey dashed lines represent the temperature cuts used to
  define cold and hot gas phases.}
\label{fig:momhist}
\end{figure}

Comparing the final configuration with the initial conditions, we find
that the cloud has increased its mass (when cooling is allowed) 
and that the corona has gained
momentum.  Quantitatively, the amount of mass and momentum transfer
depends critically on radiative cooling.  This can be seen in
Fig.~\ref{fig:momhist}, which shows the distributions of
mass (upper panel) and momentum (lower panel) as functions of
temperature after 50$\Myr$ for the standard simulation
with $v_0= 100 \kms$ and, for comparison, for its adiabatic
counterpart.  In both panels the histograms peak around the
temperatures of the coronal gas and the initial cloud temperature ($T
= 2\times10^6 \K$ and $T = 10^4 \K$, respectively), and the
distribution around these peaks is rather narrow.  From the bottom
panel of Fig.~\ref{fig:momhist} it is clear that there is an exchange
of momentum between the cloud and the corona, but the corona acquires
\textit{less} momentum when the gas is allowed to cool, as can be
inferred from the heights of the peaks at $T > 10^6\K$. This can be
better understood if we recall how the mass of the gas is distributed
among the various phases. At the beginning, most of the gas is at the
cloud and the corona initial temperatures, and only a small fraction
of the total mass is at intermediate temperatures. The momentum of the
cloud is mostly transferred to this intermediate temperature component
as a result of the mixing. Because this gas is in the temperature
range $T \simeq 1-5\times 10^5 \K$, in which the cooling function
reaches its maximum, it can cool very effectively. The consequence of
this cooling is a mass transfer from the corona towards the cold
gas. The momentum removed from the cold cloud is thus retained by the
cooling gas and never transferred to the coronal gas. If the cooling
is not present, the process of condensation of the mixed gas cannot
occur and therefore the cloud continues to transfer its momentum to
the mixed gas and thus to the corona.

\subsection{Motion of the cold gas}

\begin{figure}
\centerline{\epsfig{file=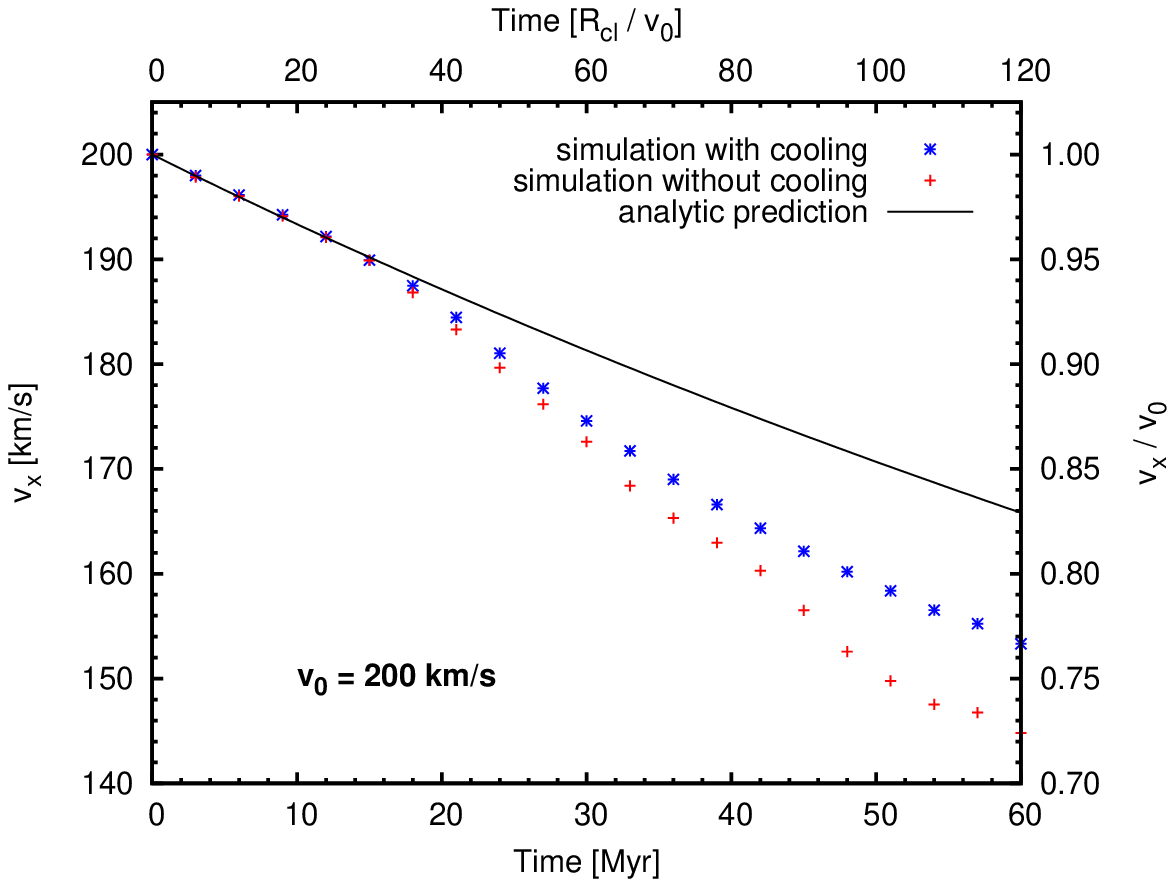,width=\hsize}}\vspace{0.3cm}
\centerline{\epsfig{file=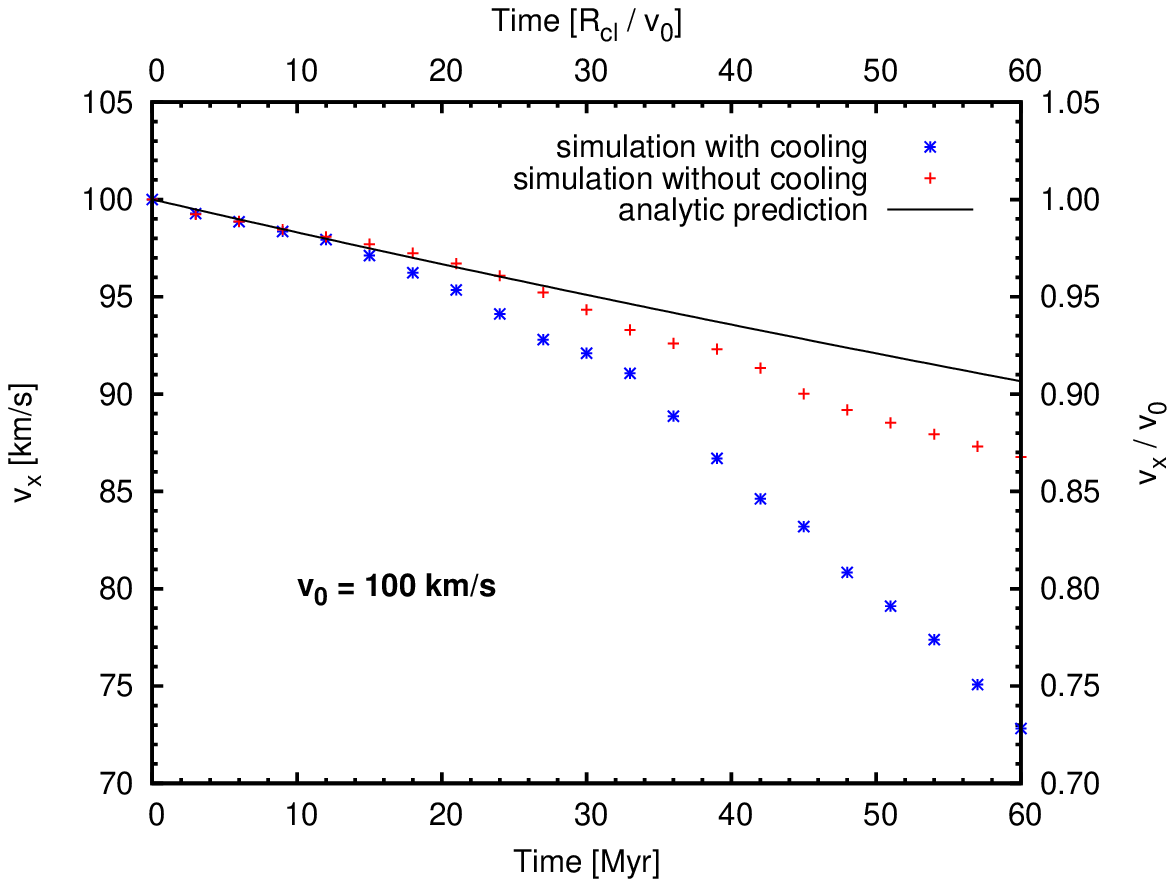,width=\hsize}}\vspace{0.3cm}
\centerline{\epsfig{file=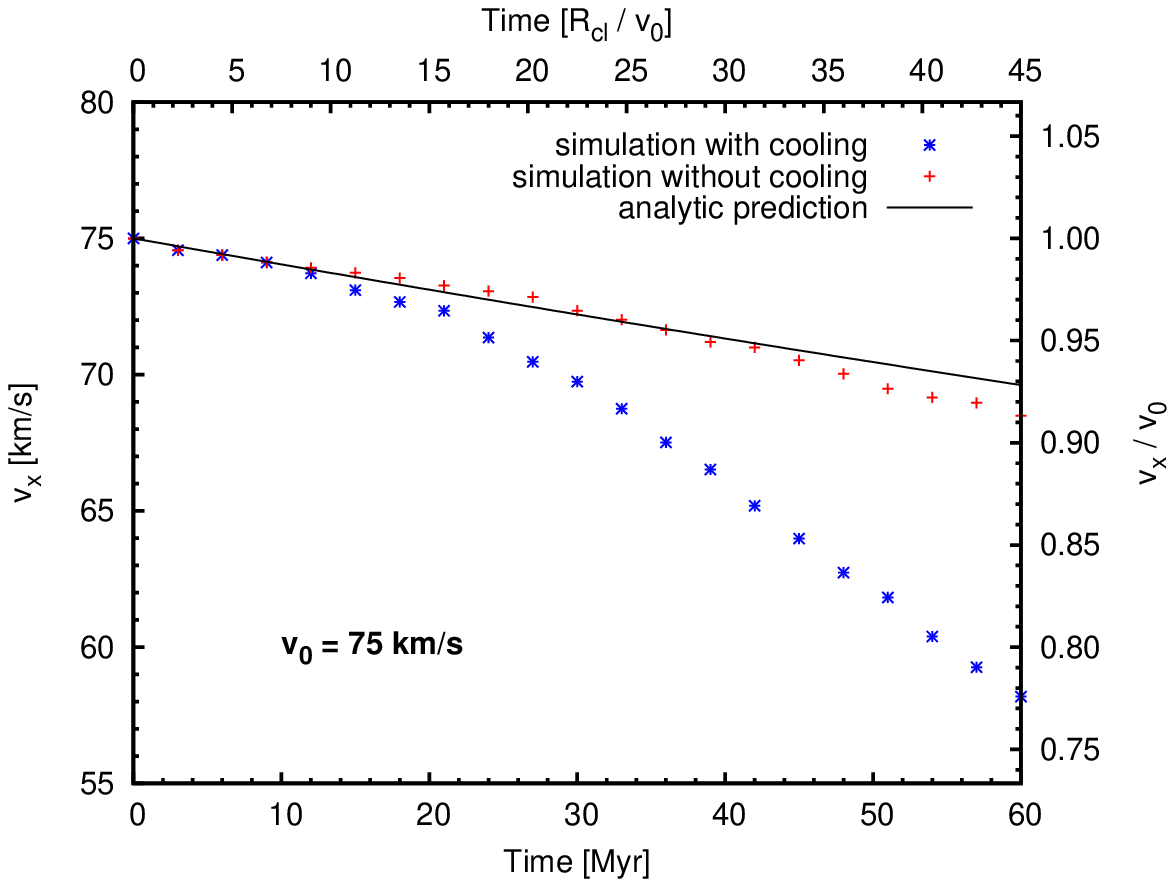,width=\hsize}}
\caption{Evolution of the centroid velocity, along the cloud's
  direction of motion, of cold gas in the dissipative (stars) and
  adiabatic (plus signs) standard simulations with $v_0 = 200 \kms$
  (top panel), $v_0 = 100 \kms$ (central panel) and $v_0 = 75 \kms$
  (bottom panel). The solid lines are the theoretical predictions in
  the case of a quadratic drag (equation~\ref{eq:drag}).}
\label{fig:cloud_centroid}
\end{figure}

In order to quantify the deceleration of the cold cloud, in the simulations
we monitor the evolution of its centroid velocity, defined as the
total momentum of the cold gas in the cloud's direction of motion $x$
divided the total mass of the cold gas.  We compare the centroid
velocity with the analytic estimate (\citetalias{FraternaliB08})
 \begin{equation}\label{eq:drag}
v(t)={v_0\over1+t/t_{\rm drag}},
\end{equation}
 where the characteristic drag time is given in terms of the cloud's initial
mass $M_{\rm cl}$, geometrical cross section $\sigma$, and coronal density
$\rho_{\rm h}$, by
 \begin{equation}\label{eq:tdrag}
t_{\rm drag}={M_{\rm cl}\over v_0\sigma\rho_{\rm h}}.
\end{equation}
In all our dissipative standard simulations, independent of the
initial speed $v_0$, after 60$\Myr$ the centroid velocity is $\sim
0.75~v_0$, significantly lower than the values predicted by
equation~(\ref{eq:drag}), which treats the cloud as a rigid body. This
discrepancy is due to a combination of cloud deformation,
hydrodynamical instabilities and condensation of cooling gas. This can be
inferred from Fig.~\ref{fig:cloud_centroid}, which shows the centroid
velocity for the dissipative (stars) and adiabatic (plus signs)
standard simulations with $v_0= 200 \kms$ (top), $v_0 = 100 \kms$
(centre) and $v_0= 75 \kms$ (bottom).
It is clear that the analytic formula (solid line in each diagram) is a
good description of the general behaviour of cold gas motion over
$\sim 60 \Myr$ only when the cooling is not allowed and the initial
relative speed is sufficiently low (see bottom panel in the figure).
When these conditions are not satisfied, the analytic formula fails
after some time, mainly because Kelvin-Helmholtz instability strips
some cold gas and the ram pressure progressively squashes the cloud
increasing its cross-section. The net effect is an enhancement of drag
effectiveness (see equation~\ref{eq:tdrag}).

Let us now compare the evolution of the centroid velocity in the presence
and in the absence of cooling.  First, it must be noticed that in
normalised units ($v_0$ for velocities and $R_{\rm cl}/v_0$ for times,
shown in the right and upper axes in the diagrams in
Fig.~\ref{fig:cloud_centroid}) the evolution of the centroid velocity
is basically the same in all the adiabatic simulations (in
the overlapping normalised time intervals). In particular, the
deviation from the analytic estimate becomes apparent at roughly the
same normalised time and it can be ascribed mainly to the flattening
of the main body of the cloud.  
On the other hand, in the presence of
cooling the deviation of the cloud speed from the analytic estimate
occurs at roughly the same {\it physical} time ($\sim 20 \Myr$) for all
explored initial speeds.
In these
dissipative cases the amount of gas that transfers from the hot to the
cold phase has an appreciable impact on the centroid velocity, so we
are no longer merely looking at the slowing of the original cloud, and
the deceleration is dominated by condensation of cooling material that
was originally at rest.

\begin{figure}
\centerline{\epsfig{file=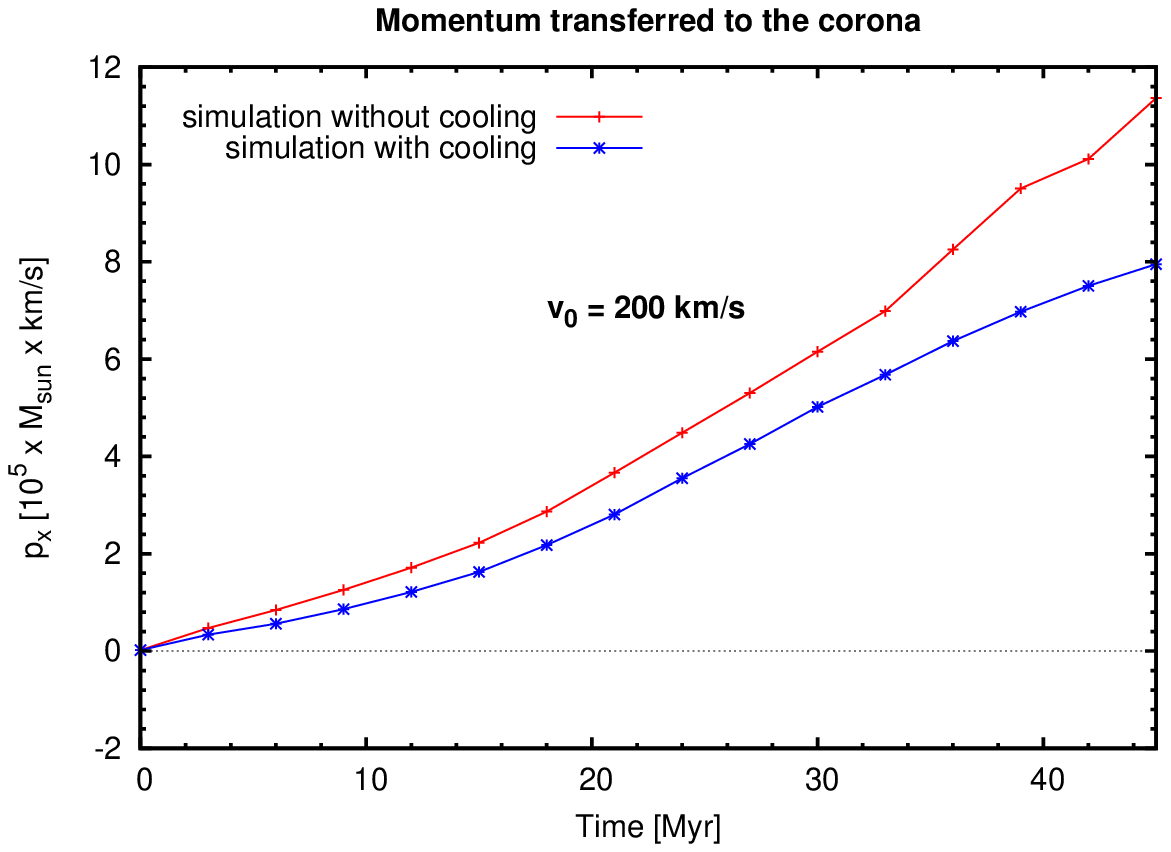,width=\hsize}}
\centerline{\epsfig{file=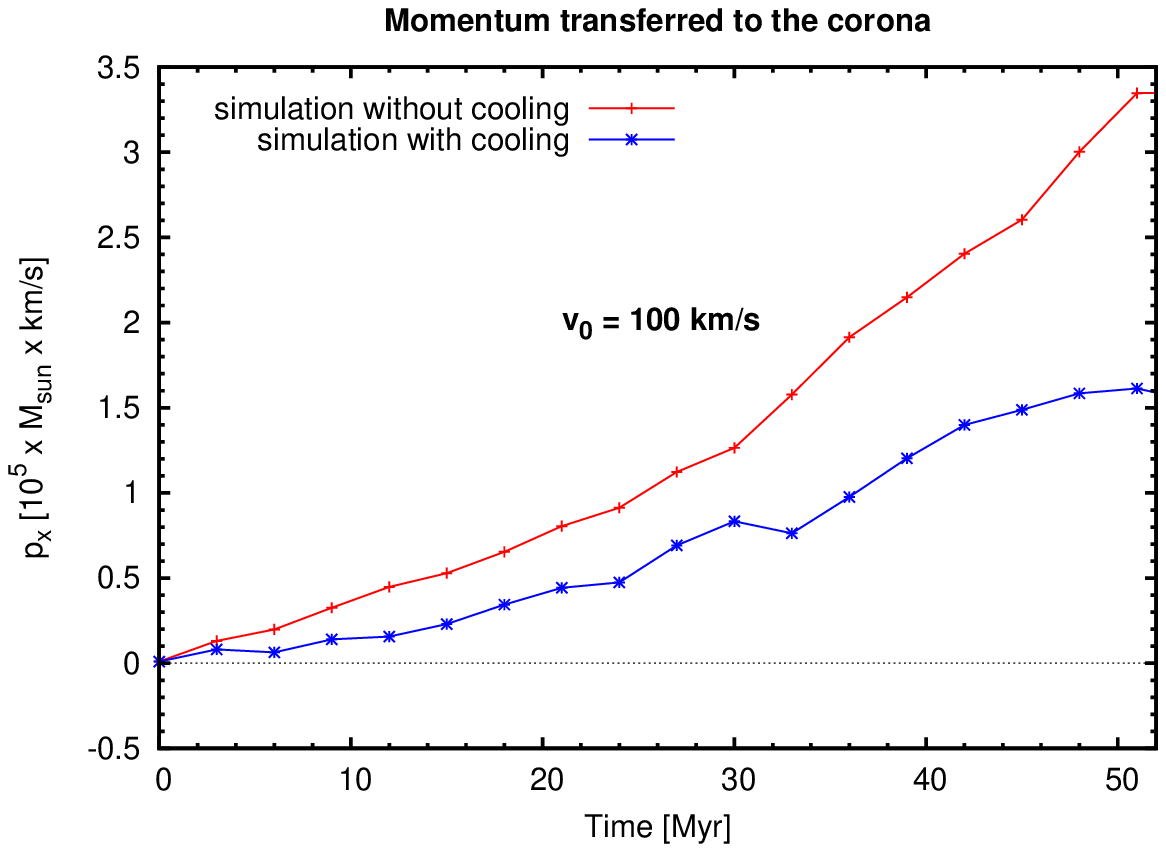,width=\hsize}}
\centerline{\epsfig{file=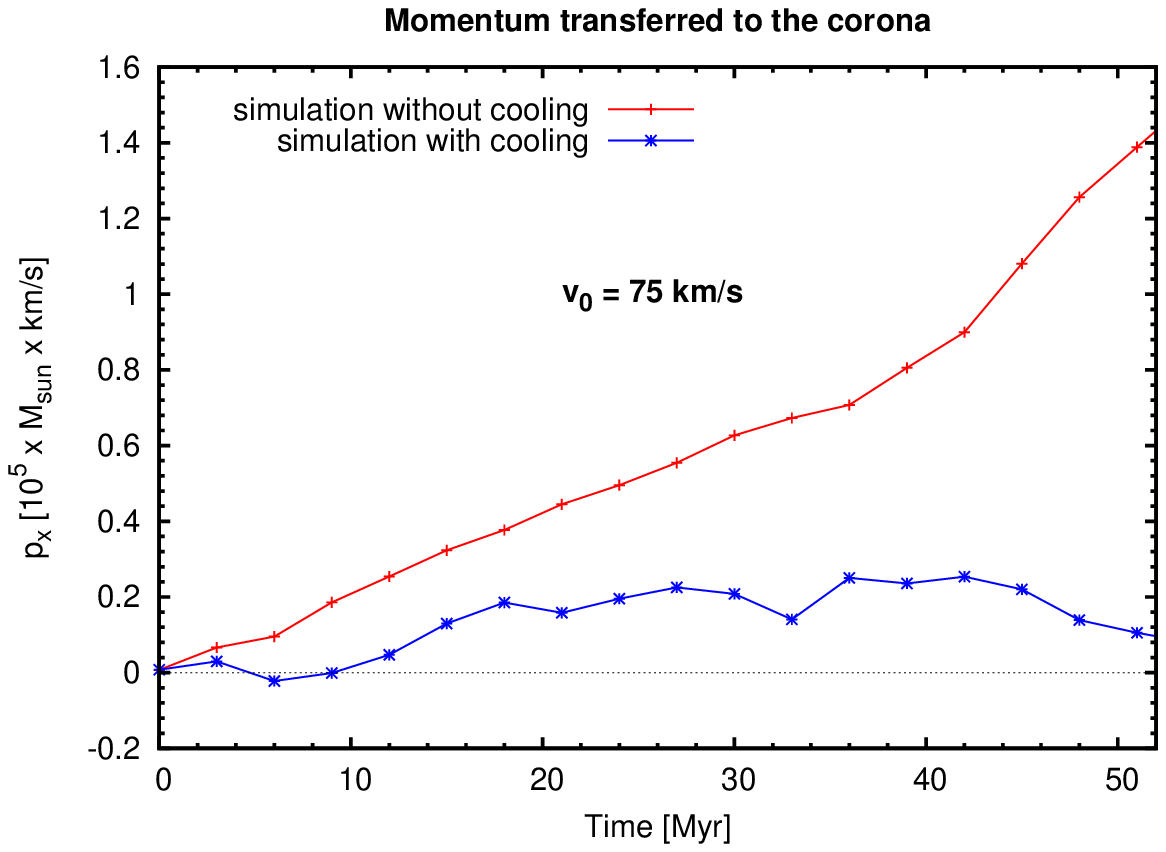,width=\hsize}}
\caption{Evolution of momentum acquired by the corona as a function of
  time in the dissipative (stars) and adiabatic (plus signs) standard
  simulations with $v_0 = 200,\,100,\,75\,\kms$ (from top to
  bottom). For comparison, we note that the total initial momentum of
  the cloud is $2.4\times10^6~(v_0/100\kms)$.
}
\label{fig:totalmom200}
\end{figure}

\subsection{Momentum of the hot gas}
\label{sect:hot_gas}

We now turn to the momentum transferred to the hot gas.  In the
simulations, the total momentum of the hot gas is computed by summing,
over the whole computational domain, the momentum of the cells
containing gas at $T > 10^6\K$.
Figure \ref{fig:totalmom200} shows the amount of momentum, along the
cloud's direction of motion $x$, acquired by the hot gas as a
function of time for the dissipative (stars) and adiabatic (plus
signs) standard simulations with $v_0=200,\,100,\,75 \kms$.  From these
diagrams it is apparent that the hot gas gains less momentum in the
presence than in the absence of radiative cooling.  The ratio between
the final coronal momentum in the dissipative and in the adiabatic
simulations is $\sim0.8$ when $v_0 = 200\kms$ (top panel), $\sim0.5$
when $v_0 = 100\kms$ (middle panel), and $\lsim 0.1$ when $v_0 =
75\kms$ (bottom panel). In particular, when $v_0 = 75\kms$ the
momentum transferred from the cold gas to the corona, after an initial
phase of growth lasting for $\sim 20\Myr$, settles to a nearly
constant value: from that moment onwards, the momentum transferred to
the corona is consistent with zero.  Thus, the simulations indicate
the existence of a \textit{velocity threshold} $v_{\rm th}$ below
which the transfer of momentum from the cold clouds to the corona is
suppressed.  Given that the cloud keeps losing momentum
(Fig.~\ref{fig:cloud_centroid}) and the total momentum is conserved,
it follows that the momentum lost by the cloud is retained by the
coronal gas, which is progressively cooling onto the
cloud's wake.

The measure of the momentum transferred from the cold to the hot gas can be
affected by the flow of coronal gas (and associated momentum) through the open
boundaries of the computational domain. This effect becomes non-negligible only
in the last $\sim 10\Myr$ of the simulations, so we take $\sim 50\Myr$ as final
time, as far as the measure of the coronal momentum is concerned (see also
Fig.~\ref{fig:momhist}). \comments{As an additional test, we also ran a
simulation with the same initial condition as the standard simulation with $v_0
= 100~\kms$, but with twice the number of grid points in the direction
perpendicular to the cloud's direction of motion (keeping the resolution fixed
and thus effectively doubling the size of the domain), finding relative
differences in the actual value of the coronal momentum within $8\%$.}

Since the density of the corona strongly influences the cooling rate
of the gas, we tested whether the velocity threshold depends on this
parameter with two additional sets of simulations with half and twice
the above coronal density. In particular we ran two simulations with
higher density ($v_0=75, 100\kms$) and three simulations with lower
density ($v_0=50,75, 100\kms$; see Table~\ref{tab:simulations}). In a
denser corona, due to the increased cooling rate, the value of the
velocity threshold rises to $\approx 85 \kms$. In the case of the less
dense corona, we found that momentum is transferred significantly also
when $v_0=75\kms$, but when $v_0=50\kms$ the momentum transfer is
consistent with zero, so $v_{\rm th}\approx 50 \kms$.  This means that
the actual value of the velocity threshold $v_{\rm th}$ is sensitive
to the coronal density, and therefore uncertain.  However, provided
that in Milky Way like galaxies the actual average coronal density 
a few kiloparsecs above the plane is in the explored interval $0.5-2\times
10^{-3} \cm^{-3}$, we can consider $v_{\rm th}\sim 50-85 \kms$ as a
fiducial range for the velocity threshold.

\section{Implications for galactic coronae and cold extra--planar gas}
\label{sect:discussion}

\begin{figure} 
\centerline{\epsfig{file=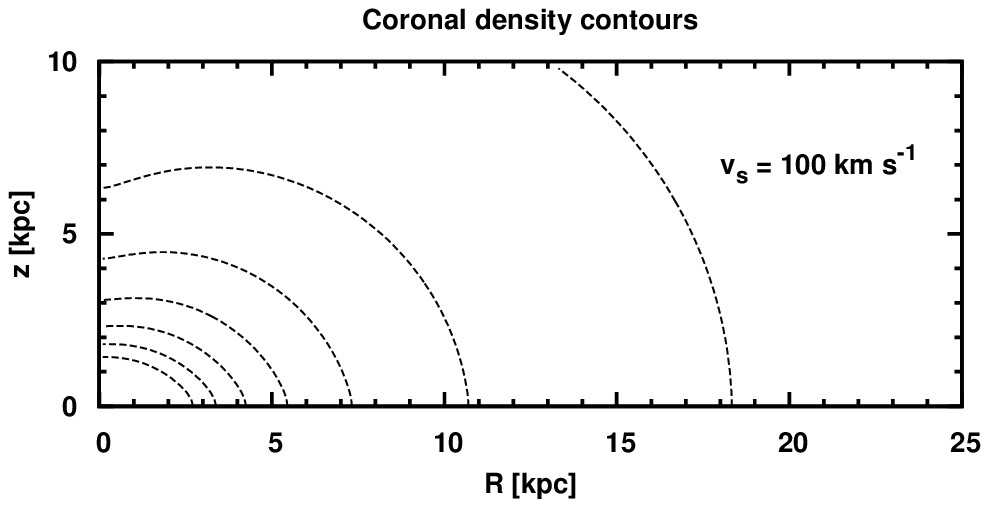,width=\hsize}}
\centerline{\epsfig{file=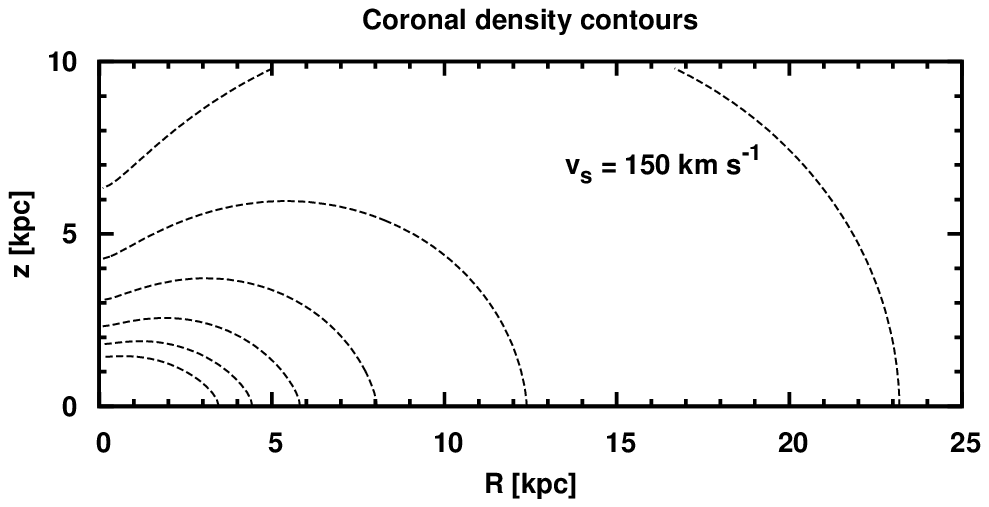,width=\hsize}}
\caption{Meridional density contours of a rotating isothermal ($T =
  2\times 10^6 \K$) corona in the Milky Way potential for two
  values of the asymptotic rotational velocity $\vs$ (see
  equation~\ref{eq:rot_cur}).  The density is normalised to the
  central ($R = z = 0$) value and the values on the contours are 0.45,
  0.4, 0.35, 0.3, 0.25, 0.2, 0.15 (the lowest contour is not visible
  in the bottom panel).  }
\label{fig:rot_coronae}
\end{figure}

\subsection{Rotating coronae}
\label{sect:corona_acc}

\citetalias{FraternaliB08} showed that if a significant part of the
momentum lost by \hi\ clouds through ram-pressure drag is absorbed by
the corona, the spin of the latter will increase on a time--scale that
is short even compared to the orbital time ($\sim100\Myr$) of an
individual cloud. If the corona did not rotate, the velocity
differences between it and individual clouds would be dominated by the
cloud's azimuthal motion, and being $\gta200\kms$ would exceed the
velocity threshold $v_{\rm th}$. Consequently, the corona would gain a
significant fraction of the momentum lost by clouds, acquiring
substantial angular momentum on a short time--scale. Hence the rotation
rate of the corona must be determined by the condition that the
relative velocity of a typical cloud and the corona is $\lta v_{\rm
  th}$. Since clouds are travelling with an average vertical velocity
$\langle v_z\rangle\sim40\kms$ \citep{Marasco10}, the rotational lag
required for this condition to be satisfied is $v_{\rm
  lag}\sim\sqrt{v_{\rm th}^2-\langle v_z\rangle^2}$. For the estimate
of $v_{\rm th} \approx 75 \kms$ given above, we obtain $v_{\rm lag}\sim
65\kms$.  Thus, we expect \coronae\ to be quite rapidly rotating.

To compute the difference in rotation velocity between the corona and
the disc at a given height, one must subtract the value for $v_{\rm
  lag}$ determined above to the value of the rotation velocity of the
\hi\ at that height. In the Milky Way, assuming a vertical gradient
for the \hi\ of $\simeq -15 \kms\kpc^{-1}$ \citep{Marasco10}, we find
that this difference is $\sim 95 \kms$ at $z \simeq 2 \kpc$,
\comments{corresponding to a coronal rotation velocity of
$\sim 125\kms$ for $v_{\sun} = 220\kms$}.

The fact that the corona is rotating, even at speeds
lower than that of the galactic disc, has important consequences for
its morphology.  
In general, the azimuthal velocity
$v_{\varphi}$ of the corona can depend on both distance from the
centre $R$ and height above the plane $z$, as is the case of
baroclinic distributions \citep[Poincar\'{e}--Wavre
  theorem;][]{Lebovitz67,Tassoul78,Barnabe06}.  The $z$ dependence
might be expected in analogy with the observed kinematics of the cold
extra--planar gas, and also as a consequence of the momentum transfer
between fountain clouds and hot gas.
\comments{By recalling the procedure outlined above, the lag of
the corona with respect to the disc should vary from 
$\sim 70$ to $120\kms$ going from $z=0.5 \kpc$ to $z=4
\kpc$, which is the typical region where fountain clouds are located.}
This corresponds to a rotational velocity of $\approx 150-100 \kms$, 
assuming again $v_{\sun} = 220\kms$. 
 However, given the large
uncertainty on the properties of the corona, here we consider much simpler
(barotropic) models of \coronae, in which
$v_{\varphi}=v_{\varphi}(R)$.  In particular, we construct isothermal
($T = 2\times10^6\K$) models of coronae in equilibrium in the
axisymmetric gravitational potential of the Milky Way, as given by
\cite{BT08}, with rotation law
\begin{equation}
v_{\varphi}(R) = \vs  \frac{R}{R + \Rs}.
\label{eq:rot_cur}  
\end{equation}
The rotation is null at the centre and raises up to the terminal value
$\vs$: the rise of the rotation velocity is steeper for smaller values
of $\Rs$.
To match the expected rotational velocity range, in Fig.~\ref{fig:rot_coronae}
the isodensity contours of two models  
with $\vs=100\kms$ (upper panel) and
$\vs=150\kms$ (lower panel) are presented. In both cases $\Rs = 1 \kpc$.
The resulting gas distributions are peanut-shaped, with axis
ratio of the isodensity surfaces in the range $0.5-0.7$.  Therefore,
if galactic \coronae\ rotate with speeds of the order of those
predicted by our model, they will appear significantly flattened when
detected in future X-ray observations. We note that this peanut-shaped
structure is also predicted in the general baroclinic case \citep{Barnabe06}.

The kinematic properties of the galactic \coronae\ are important also
for their evolution.  For instance, the thermal-stability properties
of a corona can be significantly influenced by the distribution of
specific angular momentum \citep{Nipoti10}, with implications for the
origin of the \HVC s \citep{Binney09}. However, the \HVC s are
typically thought to be relatively far from the disc, in a region not
strongly affected by disc activities, while in the present work we
have focused on the kinematics of the corona within 
a few kiloparsecs 
from the
disc, where the galactic fountain is at work. It is not clear to what
degree the kinematics of the hot gas far from the disc (which is
likely related to the formation history of the galaxy) can be
influenced by the kinematics of coronal gas near the disc.

\subsection{Vertical velocity gradient of the cold  extra--planar gas} 
\label{sect:gradient}

The simulations presented here clearly indicate that clouds of cold
gas ejected above the disc plane can lose angular momentum via
interaction with the hot medium. This mechanism appears as a natural
explanation for the observed vertical gradient in the rotation
velocity of the extra--planar gas of disc galaxies. Unfortunately, a
comparison between the results of the simulations and the observations
is not straightforward, mainly because of the idealised nature of the
simulations, in which we do not include the galaxy's gravitational
field and thus we do not follow the cloud's orbit above the
plane. However, our results can be used to obtain a rough estimate of
the expected vertical rotational gradient of the extra--planar gas,
using the following argument. Let us simplify the treatment by
assuming that a cloud, ejected with a velocity orthogonal to the disc
plane at some radius $R$, falls back onto the plane, after a time of
the order of $100\Myr$, at about the same radius $R$. Assuming a mean
velocity for the ascending part of the orbit $\langle v_z\rangle = 40
\kms$ \citep{Marasco10} and a mean maximum height above the disc for
the cold clouds $\langle z_{\rm max}\rangle = 2 \kpc$, the time to
reach the maximum height is
\begin{equation} t_{\rm orb} = \frac{\langle z_{\rm max}\rangle}{\langle v_z\rangle} \approx 50 \Myr \,. 
\end{equation}
Thus, neglecting for the moment any change in $R$ along the orbit, the
rotational-velocity gradient of the cold gas can be estimated as
\begin{equation} 
\frac{\Delta v_{\varphi}}{\Delta z} = \frac{\Delta v}{\langle z_{\rm max}\rangle} \,,
\label{eq:rot_gradient}
\end{equation} 
where $\Delta v \equiv v(t_{\rm orb}) - v_0$ is the difference between
the centroid velocity of the cold gas after $t_{\rm orb}$ and its
initial velocity.  In the hypothetical case of a static corona ($v_0 =
200 \kms$) $\Delta v \approx 40 \kms$ (see
Fig.~\ref{fig:cloud_centroid}), so
\begin{equation} 
\frac{\Delta v_{\varphi}}{\Delta z} \approx -20 \kms\kpc^{-1} \,,
\end{equation} 
while in the case of a spinning corona (such that $v_0 = 75 \kms$)
\begin{equation} 
\frac{\Delta v_{\varphi}}{\Delta z} \approx -7 \kms\kpc^{-1} \,,
\label{eq:75kmsgrad}
\end{equation}
based on the results of the dissipative simulation v75. We note that
the effect of radiative cooling is crucial for the estimate of the
gradient: using the adiabatic $v_0 = 75 \kms$ simulation we would get instead
\begin{equation} 
\frac{\Delta v_{\varphi}}{\Delta z} \approx -3 \kms\kpc^{-1}.
\end{equation} 
In the Milky Way the observed gradient is $\simeq -15\kms\kpc^{-1}$
\citep{Marasco10}, and similar values have been determined in external
nearby galaxies.  As mentioned, the above estimates neglect the
variation of $R$ along the orbit, which by conservation of angular
momentum induces a variation in
$v_{\varphi}$. \citetalias{FraternaliB08} showed that this effect
alone accounts for $\sim 50\%$ of the rotational gradient of the
extra--planar gas.  Thus, by adding this latter contribution to the
estimate in equation~(\ref{eq:75kmsgrad}), we get remarkably close to the
observed gradients.

On the basis of the above discussion we can conclude that: (i) larger
velocity gradients are expected with increasing differences in
rotational velocity between the clouds and the corona, (ii) as the
relative velocity decreases, the drag becomes an inefficient process to
decelerate the clouds, (iii) the radiative cooling of the gas has the
effect of increasing the gradient to values which are very close to
those required by the observations, and therefore (iv) the observed
gradient can be explained only by a combination of cooling and drag.
Overall, our results are consistent with the scenario developed in
\citetalias{FraternaliB08}, in which the accretion of gas by fountain
clouds is the cause of the lagging kinematics of extra--planar gas in
disc galaxies.

\section{Summary and conclusions}\label{sect:conclusions}

In this paper we studied the process of the momentum transfer between
cold clouds ejected from the disc and the corona of disc galaxies like
the Milky Way, which has a significant impact on the kinematics of
both the coronal and the cold extra--planar gas. In particular, we
presented two-dimensional hydrodynamical simulations of cloud-corona
interaction. In the more realistic simulations the gas is allowed to
cool radiatively, but, for comparison, we also considered the
corresponding adiabatic cases. The simulations are similar, but
significantly improved, with respect to those carried out by
\citetalias{Marinacci10}: here we explored a range of relative speeds
between the cloud and the hot medium ($50-200\kms$), we followed the
evolution of the system for longer time ($60\Myr$), we treated the
metallicity as a dynamical variable (assigning initially to the cloud a
metallicity ten times higher than that of the ambient gas) and we
accounted for the variation with temperature of the fluid's molecular
weight $\mu$.

From the simulations we draw the following conclusions.
\begin{enumerate}
\item There is a relative velocity threshold of about $75\kms$ between the cloud
and the ambient medium, below which the corona stops absorbing momentum. 
\item Radiative cooling is crucial for the existence of such velocity
  threshold: in adiabatic simulations the momentum transfer from the
  cloud to the hot gas is quite efficient at all explored relative
  speeds.
\item The actual value of the velocity threshold  for
Milky Way like galaxies depends on the coronal density: {\it less dense
    \coronae\ have a lower velocity threshold}. In particular, in the 
interval of coronal densities explored by our simulations $v_{\rm th} \sim
50-85\kms$.
\end{enumerate}

The process studied here has an important consequence on the
kinematics of the corona. The corona cannot be static, otherwise the
momentum transfer from the cold clouds would be very effective, but it
cannot be rotating at a speed close to the rotation speed of the disc,
because at low relative disc-corona (and thus cloud-corona) velocity
the cooling hampers any momentum transfer. In a galaxy like the Milky
Way we expect the corona to lag, in the inner regions, by $\sim 100
\kms$ with respect to the cold disc at $z\simeq2\kpc$.  Such a spinning corona
must be
characterised by a significantly flattened density distribution: based
on simple isothermal models, we estimate values of the axis ratios of
its isodensity surfaces in the range $0.5-0.7$.

Due to interaction with the corona, fountain clouds gain mass and
decelerate: based on our calculations, this effect can account for
$\approx-7 \kms\kpc^{-1}$ of the vertical gradient in rotational speed
of the extra--planar gas in disc galaxies like the Milky Way. Such a
value can be sufficient to reconcile purely ballistic fountain models
with the observed kinematics of the \hi\ extra--planar gas, giving
further support to the ``fountain plus accretion'' model of
\citetalias{FraternaliB08}.

In conclusion, we think that the presented results contribute
quantitatively to refine an emerging consistent scenario in which
galactic fountains, cold extra--planar gas and galactic \coronae\ are
strictly interlaced. This scenario makes specific predictions that
can be tested in the future with soft X-ray observations detecting the
elusive galactic \coronae\ and with improved \hi\ measures of the cold
extra--planar gas. From the theoretical point of view, the next step
will be the extension to three dimensions of the two-dimensional
calculations here presented, and ideally one would like to realise
``global'' simulations, in which the entire galaxy is modelled, with
supernova-driven fountain clouds orbiting through the hot coronal
medium under the effect of the galactic gravitational field.

\section*{Acknowledgments} 
We acknowledge the CINECA Award N. HP10CINIJ0,
2010 for the availability of high performance computing resources and support.
FM, FF, CN and LC are supported by the MIUR grant
PRIN 08.

\bibliographystyle{mn2e}
\bibliography{bibliography}{}

\label{lastpage}

\end{document}